# Hierarchical Multimodel Ensemble Estimates of Soil Water Retention with Global Coverage

Yonggen Zhang[1,2], Marcel G. Schaap[2], and Zhongwang Wei[3,4]

[1]Institute of Surface-Earth System Science, Tianjin University, Tianjin 300072, P.R. China

[2]Department of Soil, Water and Environmental Science, University of Arizona, Tucson, AZ 85721, USA

[3]School of Forestry and Environmental Studies, Yale University, New Haven, Connecticut, 06520, USA

[4]River and Environmental Engineering Laboratory, Department of Civil Engineering, University of Tokyo, Tokyo, 113-8656, Japan

Correspondence to:

Yonggen Zhang: ygzhang@tju.edu.cn

Marcel G. Schaap: mschaap@email.arizona.edu

**Key Points:**

- The performance of 13 popular models for estimating soil water retention was quantified using a dataset with global coverage.
- Relative to individual models, weighted multi-model ensembles had improved performance with the best obtained with the full ensemble.
- High-resolution global maps of soil hydraulic properties suitable for earth system modeling were produced.

**Abstract**

A correct quantification of mass and energy exchange processes among land surface and atmosphere requires an accurate description of unsaturated soil hydraulic properties. Soil pedotransfer functions (PTFs) have been widely used to predict soil hydraulic parameters. Here, 13 PTFs were grouped according to input data requirements and evaluated against a well-documented soil database with global coverage. Weighted ensembles (calibrated by four groups and the full 13-member set of PTFs) were shown to have improved performance over individual PTFs in terms of root mean square error and other model selection criteria. Global maps of soil water retention data from the ensemble models as well as their uncertainty were provided. These maps demonstrate that five PTF ensembles tend to have different estimates, especially in middle and high latitudes in the Northern Hemisphere. Our full 13-member ensemble model provides more accurate estimates than PTFs that are currently being used in earth system models.

**Plain Language Summary:** The availability of data on soil water retention is essential to quantify mass and energy exchange processes between land surface and atmosphere. In large-scale applications, soil water retention is usually estimated with statistical models that, unfortunately, were developed based on diverse databases with different methods and predictors and have unknown reliability. Using a global database we developed a method that unifies thirteen previously published models which allows for estimates of the critical soil properties under data-poor to data-rich conditions with improved estimates. Global maps of derived properties (and their uncertainties) are therefore produced at high-resolution. The maps suggest that middle and high latitudes in the Northern Hemisphere have larger variability of the estimates. The new model can provide more accurate estimates than models currently being used in earth system models.

# 1 Introduction

Soil plays a fundamental role in mass and energy exchange processes among land surfaces, groundwater, rivers, and the atmosphere (Bittelli et al., 2015; Michael & Cuenca, 1994). Quantification of surface runoff, soil-vegetation-atmosphere transfer fluxes, groundwater recharge, as well as surface energy balances and land surface temperature must therefore rely on correctly parametrized soil hydraulic properties such as soil water retention and hydraulic conductivity characteristics (Chaney et al., 2016; Montzka et al., 2017; Verhoef & Egea, 2014; Welty & Zeng, 2018; Zhao et al., 2018). The experimental determination of soil hydraulic characteristics is time-consuming, labor extensive, and especially impractical for highly heterogeneous soils in large-scale applications (Dai et al., 2013; Gent et al., 2011; Shangguan et al., 2014, 2013). Inverse estimation of soil hydraulic parameters from measurements of observable state variables such as dynamics pressure heads and/or soil water contents face similar restrictions (Man et al., 2016; Xu et al., 2017; Zha et al., 2018; Zhang et al., 2016). Instead, soil hydraulic properties are often estimated with pedotransfer functions (PTFs) which typically are empirical data-driven models that commonly utilize available soil attributes as predictors (e.g., soil texture, bulk density, and organic carbon (OC) content). Because of their utility, PTFs have become indispensable components for predicting the dynamics of moisture content in land surface models (LSMs) and global climate models (GCMs) among many other smaller-scale applications (Van Looy et al., 2017).

Over the past decades, considerable national and international efforts have resulted in many PTFs with a variety of statistical approaches (Pachepsky & Rawls, 2004; Zhang et al., 2018). These models were calibrated using data collected at local or national scales and did not necessarily represent the diversity of the global population of soils, resulting in PTFs that have been demonstrated to produce biased predictions (Dai et al., 2013; Schaap & Leij, 1998; Vereecken et al., 2016). The use of PTFs directly should therefore raise one straightforward question: what is the performance of commonly used PTFs in the context of a global-coverage dataset?

The selection of a single PTF in predictive applications may result in statistical bias, underestimation of uncertainty, and overconfidence in predictive capabilities (Neuman, 2003). To attempt to mitigate the bias and to expand the support-scale of PTFs, uniformly or variably weighted multi-model ensemble estimates can be pursued. Dai et al. (2013) used uniformly-weighted ensemble estimates to produce maps of soil hydraulic parameters for China. Utilizing the benefit of validation data, Guber et al. (2006, 2009), however, demonstrated that uniformly weighted ensembles resulted in degraded field-scale estimates.

The main thrust of the present study is to develop a strategy for optimal weighting of PTFs for multimodel estimates in the context of a soil database with global coverage. Previously, such an effort was considered to be challenging (Kishné et al., 2017; Minasny et al., 2013), not only because the calibration dataset used to develop PTFs may be a biased selection of the world population of soils, but also because soils themselves exhibit an extreme variety in hydraulic (and other) characteristics. For example, the soils in the Sahara formed under completely different climate conditions than those found in boreal zones, leading to vastly different mineral and OC contents and a resulting change in hydraulic characteristics (Davidson & Janssens, 2006). We will therefore also investigate whether one single set of optimal values would suffice to weight PTF ensemble estimates, or whether different weightings must be used for stratifications of soils according to grain-size distribution, OC content, soil order, and mean temperature.

In this study, we identified 13 widely-cited PTFs used for estimating soil water retention which were classified into four groups according to input data requirements. Such a hierarchical grouping is critical because not all input data is available in all practical use cases. The PTFs are firstly evaluated individually against a well-documented soil database with global coverage. Optimal weights for the PTFs (for all data and four stratifications thereof) are assigned by minimizing the misfit between multimodel estimates and water retention data from a global-coverage dataset. Finally, global maps of soil water retention data and its uncertainty are produced for the SoilGrids 10km dataset (Hengl et al., 2014). We anticipate that these maps will be useful for a variety of purposes. However, we are also able to estimate the uncertainty of the multimodel estimates, which will guide further research on improved global-scale PTF estimates.

**2 Material and Methods**

**2.1 PTFs and Soil Hydraulic Functions**

Summary characteristics for the 13 PTFs selected for this study appear in Supplementary Table S1; R code implementing these models appears in Supplementary Code S1. Criteria for selection were their popularity (as indicated by the number of citations listed in Table S1) and the size of the dataset used for calibration of the PTFs. PTFs that are soil-specific or do not estimate the entire water retention curve, will not be considered in this study.

There are a number of ways by which the PTFs can be grouped and distinguished. First of all, six PTFs were derived from two publications (three each from Cosby et al., 1984, and Zhang and Schaap, 2017) and represent different approaches to establish the PTFs. Secondly, owing to attempts to construct large representative databases, considerable overlap in calibration data exists among PTFs, as discussed in Supplementary Text S1. Data used for PTFs with later publication dates are likely to have also been used for "older" PTFs. Thirdly, all models estimate parameters of several water retention functions. Five PTFs estimate the parameters of Brooks and Corey (1964) water retention model or its Campbell (1974) and Clapp and Hornberger (1978) variants; eight

PTFs estimate van Genuchten (1980) parameters. Although the functional form of the retention equation is relevant, the present work will be unable to address this due to the limited number of capillary pressures available in the global-coverage dataset used for evaluation (next section).

A key distinction among PTFs is their requirements regarding predictors, which can be sorted into four groups. Group A only requires USDA soil textural class and includes Cosby0, Carsel, Clapp, and H1*w*. Group B utilizes the numerical value of the soil textural percentages as predictors and includes Cosby1, Cosby2, and H2*w*. Group C requires additional soil bulk density (Rawls, Campbell, and H3*w*), while Group D further requires soil OC content (Wösten, Weynants, and Vereecken). We refer the reader to the references for detailed descriptions of the PTFs.

## 2.2 Dataset Used for Evaluation and Ensemble Development

The National Cooperative Soil Survey Characterization Database (NCSS, 2017) is used to independently evaluate the 13 PTFs and to establish weights for the multi-model ensemble (a brief discussion of other soils databases can be found in Supplementary Text S2). After further data quality analysis (see Text S2), 49,855 records (having 118,599 water retention points) were available for further analysis. Figures 1*a* and 1*b* show the location of the selected soil samples and the textural distribution in USDA soil textural triangle.

## 2.3 Evaluation Criteria, Multi-model Ensemble Predictions, and Bootstrap Resampling

The criteria used to evaluate different PTFs is to use root mean square error (*RMSE*) of moisture content, defined as

$$RMSE = \sqrt{\frac{1}{N_d}\sum_{i=1}^{N_d}\left(\theta_i(\psi_i) - \theta'_i(\psi_i)\right)^2} \qquad (1)$$

and $\theta'_i(\psi_i) = f\left(M_j(\mathbf{D}), \psi_i\right)$ \qquad (2)

where $\theta_i(\psi_i)$ and $\theta'_i(\psi_i)$ are measured and estimated moisture content at pressure head $\psi_i$; $N_d$ is the number of measurements for moisture content, equal to 118,599 for the entire dataset, but less for each stratification of NCSS dataset (i.e., USDA textural classes, soil OC content, soil orders, and soil temperature). **D** is a vector of predictors specific for PTF model $M_j$; $f$ is a water retention function as discussed previously, evaluated at pressure head $\psi_i$. Model selection criteria, such as *AIC* (Akaike, 1974) and *AICc* (Hurvich & Tsai, 1989) as defined in Supplementary Text S3, is used to rank and compare individual PTFs and ensemble models. Lower *AIC* or *AICc* value indicate the the preferred model.

The weights of multi-model ensemble simulation are determined by minimizing the following

$$\chi^2(\mathbf{a}) = \bar{\boldsymbol{\varepsilon}}(\bar{\boldsymbol{\varepsilon}})^T \qquad (3)$$

$$\text{where } \bar{\boldsymbol{\varepsilon}} = \frac{1}{\sum_{j=1}^{N_m} a_j} \sum_{j=1}^{N_m} a_j (\boldsymbol{\theta}' - \boldsymbol{\theta})^2 \qquad (4)$$

and *a* is a vector of weights, $a_j$, for corresponding PTF model $M_j$ ($j = 1...N_m$), $0 < a_j < 1$; $N_m$ is the number of PTF ensembles (13 when all PTFs are considered, but 3 or 4 for PTF Groups A through D); $\boldsymbol{\theta}$ and $\boldsymbol{\theta}'$ are vectors of measured and estimated moisture content with the length of calibration samples selected by the bootstrap re-sampling process (see below). Different algorithms were evaluated to minimize (3) and we found that a genetic algorithm (Scrucca, 2013, implemented in the statistical package R, version 3.4.4, Venables & Smith, 2003) was the most effective method to derive *a*.

The optimization of the ensemble weighting vectors for the entire dataset and the four stratifications was coupled with bootstrap re-sampling (Efron & Tibshirani, 1993) to obtain the uncertainty of *a*. The replica datasets that were generated further enabled us to consider error metrics for independent calibration and validation data (Zhang & Schaap, 2017). We found that 100 bootstrap replica datasets sufficed to produce stable weighting vectors.

**3 Results and Discussion**

**3.1 Evaluation of Individual PTFs**

When considering the performance of the 13 PTFs individually, we found that *RMSE* is the largest (0.0987 cm$^3$/cm$^3$) for the Carsel PTF and the lowest (0.0555 cm$^3$/cm$^3$) for the Wosten PTF (Table 1). Group-level performance is expected to improve when more predictors are used, though this is not always the case. For example, the Vereecken PTF (Group D) has an *RMSE* of 0.0658 cm$^3$/cm$^3$ whereas the simple class PTF of Cosby (Group A) has a lower *RMSE* (0.0624 cm$^3$/cm$^3$).

The notable outlier in Table 1 is the Carsel PTF, which has the worst performance but is also one of the most widely cited in the literature (Table S1). The poor performance of this model is likely due to its heuristic transformation of Brooks and Corey (1964) retention parameters into van Genuchten (1980) parameters. The Carsel PTF is based on the Rawls PTF (see Carsel and Parrish, 1988), which has much better performance (0.0629 versus 0.0987 cm$^3$/cm$^3$). *AIC* and *AICc* criteria yield results (shown in Supplementary Table S2) similar to those shown by *RMSE* for individual PTFs, and therefore will not be discussed here.

**3.2 Group Ensembles**

Summary results for the optimizations of the Group A through D ensembles as well as those for all 13-PTF models appear in Table 1. Detailed results for each replica and ensemble member are available in Supplementary Tables S3-S7. The supplemental results exhibit small differences in *RMSE* values for the calibration and validation members of each bootstrap replica indicating that each ensemble model is stable. The standard deviation of weights for individual PTFs in each ensemble was less than 0.02; the sum of ensemble weights was always equal to 1.

*RMSE* values of ensemble models are all smaller than those of individual ensemble members, decreasing monotonically from 0.0624 (Group A) to 0.0528 (Group D), declining further to 0.0517 cm$^3$/cm$^3$ (full 13-member ensemble). It is worth noting that there is a substantial improvement from Group B to C (0.0053 cm$^3$/cm$^3$), but only a small improvement from Group C to D (0.0008 cm$^3$/cm$^3$). Bulk density provides information about total porosity but is negatively correlated with soil OC content (Heuscher et al., 2005; Zacharias & Wessolek, 2007). Both therefore provide similar information (Périé & Ouimet, 2008; Prévost, 2004), while the Group C to D improvement is statistically meaningful (see corresponding *AIC* and *AICc* values in Table S2) and suggests that including OC provides additional information. It is noted that the presented weights are based on all available retention data in selected NCSS dataset, while optimization using different subsets of retention points yields different weighting for different PTFs (See Text S4 and Figure S1).

The Clapp PTF is the dominant member in the Group A ensemble (with a weight of 0.6067, see Table 1), Cosby1 (0.5961) in Group B, Rosetta-H3*w* (0.5529) in Group C, and Weynants (0.5422) in Group D. The weighting values of individual PTF change in both absolute and proportional terms for the ensemble of all PTFs. In this case, the Rosetta-H3*w*, Wosten, and Weynants PTFs carry roughly equal weights (0.17 to 0.20), while the Clapp PTF has a surprisingly large contribution (0.14), given its comparatively large individual *RMSE* and its simple USDA textural class input requirements. The remaining 30% of the ensemble weights were carried by the nine other PTFs. The results suggest that there is merit in pursuing multi-model ensemble predictions, rather than just picking the individual PTF with the best performance. In addition, the full 13-member ensemble should be considered if possible.

**3.3 Optimization of Weights for Different Soil Characteristics**

The previous section provides optimal weights for the 13 selected weights given the 49,855 selected samples in the NCSS database. There is no guarantee that this subset of data represents the actual distribution of soils in the world, while there is even less of a guarantee that the original calibration data used to establish the 13 PTFs are representative. By stratifying the selected NCSS data by soil textural class, soil OC content, soil order and mean soil temperature, it is possible to evaluate whether better estimates can be made by re-optimizing *a* in (3) when the data is stratified according to these variables. Figures 1*c*, 1*d*, and 1*e* show the maps of soil OC content, soil temperature, and soil orders within the contiguous USA. Data from other continents was included in the analysis but is not shown for clarity; soil texture did not exhibit visually useful patterns and therefore is not displayed in Figure 1.

Summary results shown in Figure 2 (with detailed bootstrap-level data appearing in Supplementary Tables S8-S11) indicate that PTF weights do indeed change for the different stratifications of the data. When stratified by USDA textural class (Figure 2*a*), the weight for all PTFs vary substantially (except Cosby0 and Carsel which remain low), which we believe is caused by data-selection effects when the original PTFs were calibrated. This causes some PTFs to outperform others for some parts of the textural triangle, but become inferior elsewhere. Different weights vectors are also obtained when the data are stratified by soil OC content (Figure 2*b*) or soil order (Figure 2*c*). When the data are stratified by taxonomic temperature regime (Figure 2*d*), the PTF weights exhibit less variation.

The results in Figure 2 indicate the likelihood that better estimates can be obtained when the weights are determined for different stratifications of the data. When these weights are used at face value, the *RMSE* is reduced from 0.0517 to 0.0511, 0.0511, 0.0506, and 0.0490 cm$^3$/cm$^3$ for stratifications according to textural class, OC content, soil order, and taxonomic temperature, respectively. We note here that the stratifications by soil order and temperature regime could only be conducted for a reduced number of samples (91,303 and 80,223 samples, respectively). Strictly speaking, the *RMSE* values for taxonomy and temperature cannot be compared accurately to that of the 13-member ensemble and we therefore resort to *AIC* and *AICc* values in Table S2, which confirm that the improvements by re-optimizing the PTF weights for different stratifications of the data are statistically significant.

**3.4 Prediction of Field Capacity and Wilting Point from Multi-model Ensembles**

The ensemble models displayed in Table 1 were applied to the SoilGrids 10km dataset of Hengl et al. (2014) into moisture contents at saturation (0 cm pressure head), field capacity (330 cm) and wilting point (15,000 cm). These derived quantities are useful metrics because saturated water content is sometimes considered to be equal to the porosity, while field capacity is accepted to be the pressure where gravitationally induced drainage of water is minimal, and wilting point is considered as the pressure head where most vegetation ceases to extract water from soils (Dane & Topp, 2002; Jury & Horton, 2004; Klute, 1986). The difference between saturation and field capacity has previously been used to estimate saturated hydraulic conductivity (Ahuja et al., 1989) and provides a path forward to also estimate unsaturated hydraulic conductivity (van Genuchten, 1980; Mualem, 1976). The SoilGrids dataset was used because it provides accurate, high-resolution global maps of soil texture, bulk density, and soil OC content derived by automated soil mapping (Luo et al., 2016; Montzka et al., 2017). Although the 10 km resolution version was used here, maps at finer (1 km or 250 m, see Hengl et al., 2017) or coarser (Montzka et al., 2017) resolutions can also be generated. We show here the mean values and corresponding coefficient of variation (CV) of the 13-member ensemble estimates in Figures 3. Maps of ensemble estimates and CV values of saturated water content, field capacity and wilting point for Groups A through D and the 13-member ensemble estimates can be found in Figures S2-S4. The CV values represent the ratio of the standard deviation to the mean of the 100-member bootstrap estimates for each grid location.

Figure 3 shows that low saturated water content values are mainly located in most of the Southern Hemisphere and low- and mid-latitudes of the Northern Hemisphere, while high values are found in high latitudes of the Northern Hemisphere. Low field capacity and wilting points are found in the Sahara and Arabian Peninsula, while high field capacity values are shown in Canada, most of Siberia, and part of South and Southeast Asia. High wilting point values are found in Mexico, Central America, South and Southeast Asia, and parts of South America and parts of Tropical Africa, as well as locations in Canada and Siberia. Figures S2 and S3 respectively show that, compared to Group A and B, Group C and D ensembles exhibit substantially higher saturated water contents and field capacities in parts of Canada and Siberia; these higher estimates persist in the 13-member ensemble (Figures S2*i* and S3*i*). This is presumably due to the information provided by bulk density and organic carbon content, used by Groups C and D ensembles. Soils in these regions are known to have high OC contents which are inversely correlated with bulk density (Zacharias & Wessolek, 2007). Given that *RMSEs* produced by Groups C and D ensembles are lower than those of Groups A and B, it is likely that PTFs in Groups A and B systematically underpredict saturated water content and field capacities for the northern regions. There are also distinct differences between the maps for wilting point produced by the Groups A, B, C ensembles and those generated with the Group D and the full 13-member ensemble (Figure S4). This implies that soil organic carbon content in PTF models strongly affects estimates of the wilting point.

The choice of PTF and PTF-ensemble may therefore lead to an inherent bias in moisture content estimates for particular geographical regions. For example, De Lannoy et al. (2014) found significantly improved soil moisture simulation of catchment LSMs compared with observations in four watersheds in the US by utilizing the Group D, Wosten PTF, instead of the Cosby0 in Group A PTF. The variation of CV values are discussed in Supplementary Text S5. High values of CV might serve as a guide where PTF improvement should be conducted.

**4 Summary and Conclusions**

Our work leads to five major points:

1. Thirteen widely-used pedotransfer functions (PTFs) for the estimation of water retention characteristics were grouped in four classes according to input data requirements and evaluated on independently-measured moisture contents available in a large soil dataset with global coverage (NCSS dataset).

2. Weighted multimodel ensemble estimates for each PTF group resulted in improved performance. A further improvement was achieved by a weighted ensemble of all 13 models. In this case four of the models carried 70% of the weights, while the remaining nine models carried only 30% of the weights.

3. Model weights changed for stratifications according to USDA textural class, OC content, soil order, and soil temperature classification, which indicates that each of the PTFs had intrinsic biases. Improved PTFs can be obtained when the weights are determined for different stratifications of the data, and further research remains necessary to derive a meta PTF that effectively deals with soil-dependent weights.

4. Maps of field capacity and wilting point (and their uncertainties) were derived for the ensemble models. The maps produced in the present study have a resolution of 10 km; higher resolution maps and data structures with complete retention curves (and associated uncertainty) can also be generated. These maps demonstrated that the calibrated PTF ensembles tend to have different estimates, especially in mid and high latitudes in the Northern Hemisphere.

5. The ensemble PTFs provide estimates that are superior over PTFs currently being used in earth system modelling. Use of estimates by weighted PTF ensembles may therefore provide more accurate estimates of moisture content for soil water balance models, hydrological and ecological models, crop growth models, land surface models, weather forecast models, air quality models, and global climate models.

**Acknowledgments**

This study is supported by the National Natural Science Foundation of China (grant number: 41807181). We thank Shlomo Neuman for the derivation of model selection criteria and anonymous reviewers for their constructive comments, which improved the quality of the manuscript. The global maps (in GeoTIFF format) of mean values and coefficients of variation of saturated water content, field capacity, and permanent wilting point in 10 km resolution estimated from overall 13-PTF ensemble model can be downloaded from: http://www.u.arizona.edu/~ygzhang/download.html.

Rawls, W. J., & Brakensiek, D. L. (1985). Prediction of soil water properties for hydrologic modeling. In *Watershed Management in the Eighties* (pp. 293–299). American Society of Civil Engineers.

Schaap, M. G., & Leij, F. J. (1998). Database-related accuracy and uncertainty of pedotransfer functions. *Soil Science*, *163*, 765–779. https://doi.org/10.1097/00010694-199810000-00001

Scrucca, L. (2013). GA: A Package for Genetic Algorithms in R. *Journal of Statistical Software*, *53*(4), 1–37. https://doi.org/10.1359/JBMR.0301229

Seber, G. A. F., & Lee, A. J. (2003). *Linear Regression Analysis* (Second Edi). Hoboken, NJ, USA: John Wiley & Sons, Inc.,. Retrieved from http://dx.doi.org/10.1002/9780471722199.ch3

Seber, G. A. F., & Wild, C. J. (1989). *Nonlinear Regression*. *StatBook*. Hoboken, NJ, USA: John Wiley & Sons, Inc.,. Retrieved from http://doi.wiley.com/10.1002/0471725315

Shangguan, W., Dai, Y., Liu, B., Zhu, A., Duan, Q., Wu, L., et al. (2013). A China data set of soil properties for land surface modeling. *Journal of Advances in Modeling Earth Systems*, *5*(2), 212–224. https://doi.org/10.1002/jame.20026

Shangguan, W., Dai, Y., Duan, Q., Liu, B., & Yuan, H. (2014). A global soil data set for earth system modeling. *Journal of Advances in Modeling Earth Systems*, *6*(1), 249–263. https://doi.org/10.1002/2013ms000293

Van Engelen, V. W. P. (2011). Standardizing soil data (e-SOTER regional pilot platform as EU contribution to a Global Soil Information System). In *International Innovation*. June, 48-49.

van Genuchten, M. T. (1980). A Closed-form Equation for Predicting the Hydraulic Conductivity of Unsaturated Soils. *Soil Science Society of America Journal*, *44*(5), 892–898. https://doi.org/10.2136/sssaj1980.03615995004400050002x
20

# Tables

Table 1. Root mean square error (*RMSE*, unit: cm$^3$/cm$^3$, bold font) of moisture content of individual PTF and ensemble models for different groups and all PTFs optimized based on NCSS dataset ($n$ = 118,599; optimization are mean values based on 100 bootstrap replicas); corresponding mean weights of the ensemble models are also shown.

| Group* (required input data) | Group A (USDA texture class) | | | | Group B (USDA texture percentage: sand, silt, and clay) | | | Group C (as B, but with bulk density) | | | Group D (as, C but with soil organic carbon) | | |
|---|---|---|---|---|---|---|---|---|---|---|---|---|---|
| | Cosby0 | Carsel | Clapp | Rosetta3-H1$w$ | Cosby1 | Cosby2 | Rosetta3-H2$w$ | Rawls | Campbell | Rosetta3-H3$w$ | Wosten | Weynants | Vereecken |
| *RMSE* of individual model | **0.0624** | **0.0987** | **0.0627** | **0.0681** | **0.0607** | **0.0620** | **0.0628** | **0.0629** | **0.0675** | **0.0589** | **0.0565** | **0.0555** | **0.0658** |
| Weights of group ensemble model | 0.1509 | 0.0936 | 0.6067 | 0.1487 | 0.5961 | 0.0028 | 0.4010 | 0.1056 | 0.3415 | 0.5529 | 0.4565 | 0.5422 | 0.0013 |
| *RMSE* of group ensemble model | **0.0601** | | | | **0.0589** | | | **0.0536** | | | **0.0528** | | |
| Weights of overall ensemble model | 0.0330 | 0.0106 | 0.1437 | 0.0250 | 0.0257 | 0.0264 | 0.0414 | 0.0325 | 0.0858 | 0.1704 | 0.1980 | 0.1854 | 0.0223 |
| *RMSE* of overall ensemble model | **0.0517** | | | | | | | | | | | | |

* Different Groups (A, B, C, and D) are classified based on input data. See text and Supplementary Table S1 for detailed explanations of the evaluated PTFs.

**Figures**

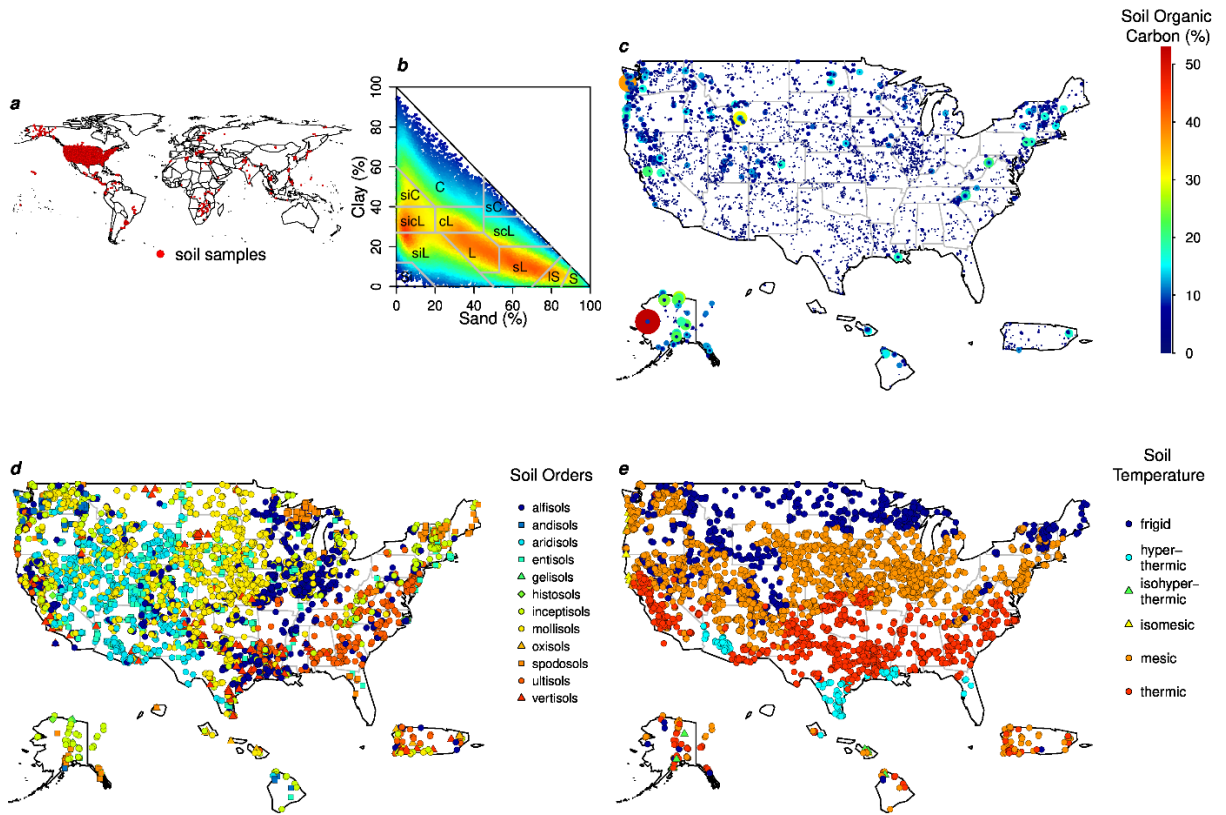

**Figure 1. Map of NCSS soil samples according to different stratifications.** (*a*) Map of soil samples in the world. (*b*) NCSS soil samples in USDA soil textural triangle. Red color represents high density, blue low density. (*c*) Map of soil organic carbon (OC) content within US territory. Red color and large circles represent high organic carbon, while blue and small circles indicate low organic carbon. (*d*) Map of soil orders and (*e*) soil temperature.

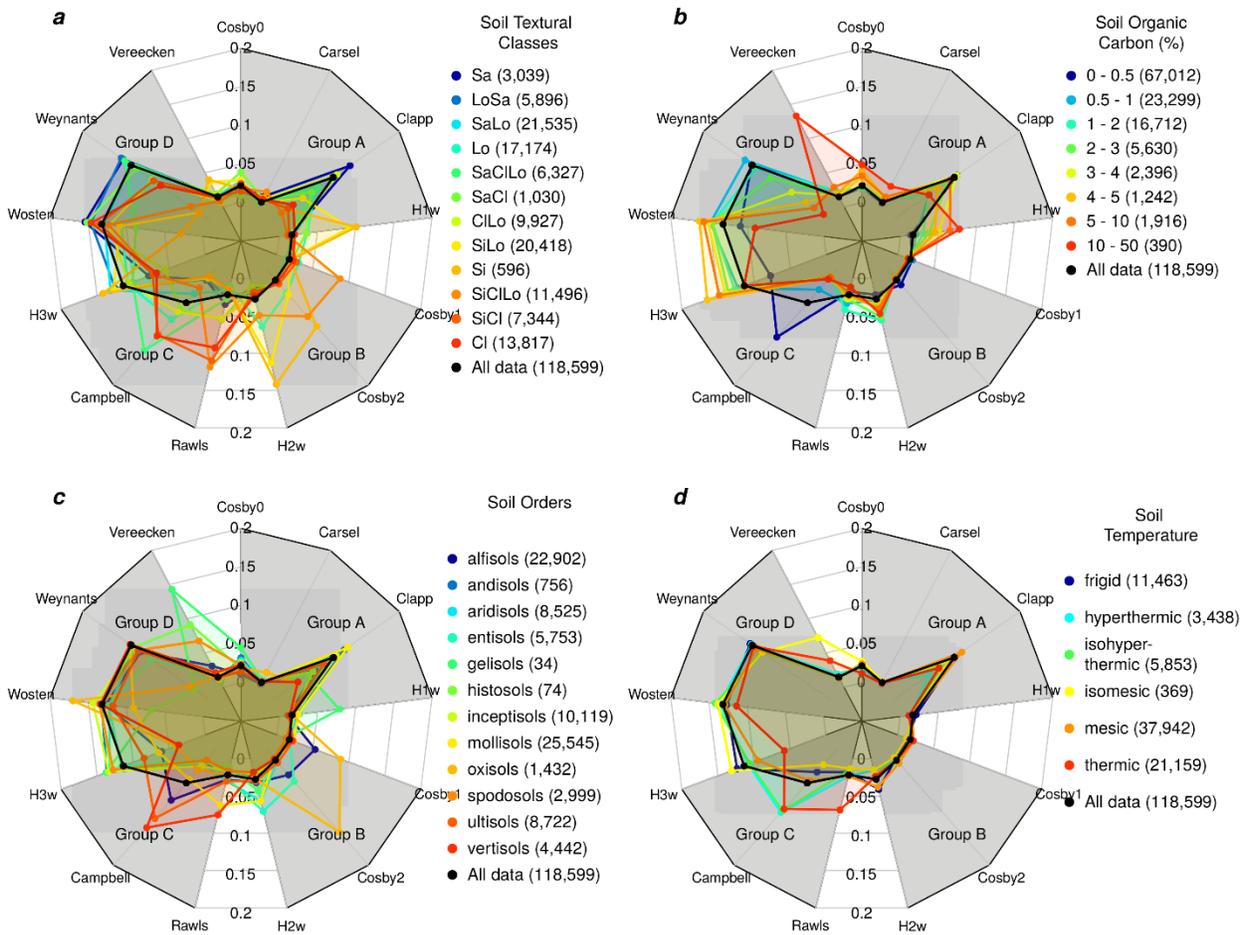

**Figure 2. Radar plots showing the weights for different PTFs in optimized ensemble average models (sum of weights for all 13 ensemble PTFs equals to one).** Optimization based on (*a*) different USDA soil textural classes, (*b*) soil organic carbon, (*c*) soil orders, and (*d*) soil temperature. Number in the legend indicates the number of retention points in each stratification. Weights optimized based on all samples are also shown as black dots.

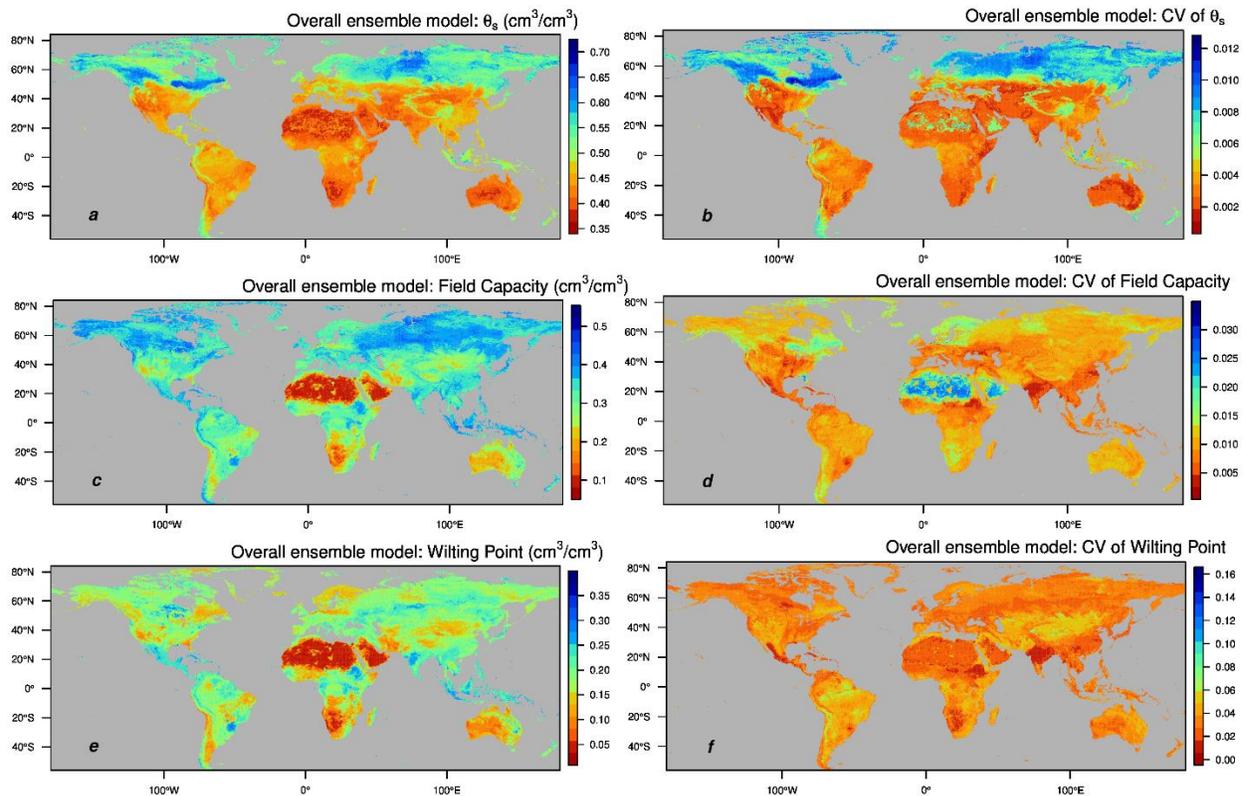

**Figure 3. Global maps of mean values and coefficients of variation (CV) of saturated water content ($h$ = cm; *a* and *b*), field capacity ($h$ = 330 cm; *c* and *d*) and permanent wilting point ($h$ = 15,000 cm; *e* and *f*) in 10 km resolution from overall 13-PTF ensemble model**. Weights for different PTFs are based on the average values of 100 bootstrap replicas of optimized weights. Calculations are based on the surface soil of SoilGrids 10 km data set (Hengl et al., 2014).

**Supporting Information**

**Text S1 Description of Overlap among Datasets used for development of individual PTFs**

Not all PTFs have independent data sets. Data used to establish the *Rawls* (Rawls and Brakensiek, 1985) PTF was used by *Carsel* (Carsel and Parrish, 1988) and partially for the *Rosetta3* PTFs published by Zhang and Schaap (2017). The *Cosby0*, *Cosby1*, and *Cosby2* PTFs (in Cosby et al., 1984) all use the same data, while *Vereecken* (Vereecken et al.,1989) data was used for *Weynants* (Weynants et al., 2009) and included in the database used in *Rosetta3* and was included in the European database used by the Wösten et al. (1999) PTF.

**Text S2 Other Potential Database and NCSS Data Quality Analysis**

WoSIS dataset (Batjes et al., 2017) has a wider global support than the NCSS (NCSS, 2017) database that was selected for the present study. Data available in WoSIS is derived from a large number of sources, including the NCSS database (NCSS, 2017) and Africa

Soil Profiles database (Leenaars, 2013), ISRIC Soil Information System (Batjes, 2009), and Soil and Terrain databases(Van Engelen, 2011).

A preliminary analysis of the WoSIS data, however, indicated that bulk density was not available for many soils records, which would have precluded the evaluation of group C and D PTFs, such as Rawls and Brakensiek (1985), Campbell and Shiozawa (1992), Vereecken et al. (1989), Wösten et al. (1999), and Zhang and Schaap (2017). The vast majority of the WoSIS records that had valid bulk density values were derived from the NCSS database. Because the WoSIS sources may further have used different measurement protocols to acquire the data that could lead to systematic source-dependent differences among data, we decided to limit the present study to NCSS data only, which was acquired under uniform analysis procedures (NCSS, 2017). Although most of the selected NCSS data is from the contiguous USA, 5,031 samples were derived from Alaska, Hawaii, South America, Africa, Europe and Asia; in addition 5,209 of the 49,855 selected NCSS samples do not have specified coordinates, but are presumably from the contiguous USA. We also evaluated the Food and Agriculture Organization of the United Nations (FAO) soil profile dataset, but determined that this data could not be used for the present analysis.

The major merit of the NCSS data is their well-documented and consistent analysis procedures. This minimizes systematic artifacts in the evaluation data. The database includes 63,565 soil pedons with multiple soil horizons, resulting in a total of 397,212 records with sometimes sparse data. In order to evaluate all 13 PTFs, we required that soil texture data, bulk density, soil OC content must be available. Also, each record must at least have one measured moisture content (measured at 60, 100, 330, 1000, 2000, or 15,000 cm pressure head).

Moisture contents were determined in the gravimetric method and were converted to volumetric values by multiplying corresponding bulk density at 330 cm pressure head. Some "outlier" soil samples were removed, such as those with soil bulk densities outside the range of [0.5, 2.0] g/cm$^3$, soil moisture content at 330 cm pressure head lower than that at 15,000 cm. In a few rare cases, moisture content higher than 1 cm$^3$ cm$^{-3}$, were found in the database which are presumably due to data entry errors. To prevent unseasonable results, we removed all volumetric moisture contents larger than 0.6 cm$^3$ cm$^{-3}$ at 330 cm and 15,000 cm pressure head.

**Text S3 Model Selection Criteria**

Consider a set of $k$ alternative models, $M_k$, $k = 1, 2,…K$. The Gaussian likelihood function is:

$$L(\hat{\boldsymbol{\beta}}_k | \mathbf{z}^*) = (2\pi)^{-N_z/2} \left|\boldsymbol{\sigma}^2 \boldsymbol{\omega}^{-1}\right|^{-1/2} \exp(-\frac{1}{2} \frac{\boldsymbol{\varepsilon}^{*T} \boldsymbol{\omega} \boldsymbol{\varepsilon}^*}{\sigma^2}) \tag{A1}$$

where $\hat{\boldsymbol{\beta}}_k$ is the maximum likelihood estimate of a vector $\boldsymbol{\beta}_k$ of $N_k$ adjustable parameters associated with model $M_k$; $\mathbf{z}^*$ is an observed vector of $N_z$ ($N_z$ = 118,599 is the number of water content in this study stated in section 2.2) system state variables $\mathbf{z}$ in space-time; $\boldsymbol{\sigma}^2$ is a vector of known or unknown nominal error variance; $\boldsymbol{\omega}^{-1}$ is a known weight matrix; $\boldsymbol{\varepsilon}^* = \boldsymbol{\theta} - \boldsymbol{\theta}'$ is a vector of the differences between observed and simulated water content. By taking logarithmic on both sides of (A1), it can be rewritten as

$$-2\ln[L(\hat{\boldsymbol{\beta}}_k \mid \mathbf{z}^*)] = N_z \ln(2\pi) + N_z \ln \sigma^2 + \ln|\boldsymbol{\omega}^{-1}| + \frac{\boldsymbol{\varepsilon}^{*T}\boldsymbol{\omega}\boldsymbol{\varepsilon}^*}{\sigma^2} \tag{A2}$$

Let us use $J = \boldsymbol{\varepsilon}^{*T}\boldsymbol{\omega}\boldsymbol{\varepsilon}^*$ (A3)

(A3) is the least square fit of computed and measured water contents in this study, which has a relationship with *RMSE* of Equation (1) in the main text as

$$J = N_z \times RMSE^2 \tag{A4}$$

$J^*$ is then defined as the arithmetic mean value of $J$ when $J$ becomes approximately stable, independent of the PTFs. In this study, $J^*$ is calculated as the average of the values of $J$ associated with all 13 PTF models because of reasonable range of all PTF estimates. It is calculated that $J^* = 513.27$.

Since $\sigma^2$ is often difficult to evaluate a prior, and it is unknown in this study. Therefore, the estimated error variance $\hat{\sigma}^2$, calculated as:

$$\hat{\sigma}^2 = \frac{J^*}{N_z} \tag{A5}$$

is used to estimate $\sigma^2$ (*Carrera and Neuman*, 1986; *Seber and Wild*, 1989; *Seber and Lee*, 2003; *Ye et al.*, 2008). We obtain $\hat{\sigma}^2 = 0.004329$, and we can then calculate the value of $\frac{\boldsymbol{\varepsilon}^{*T}\boldsymbol{\omega}\boldsymbol{\varepsilon}^*}{\hat{\sigma}^2}$ for different models.

In (A2), $N_z \ln(2\pi)$, $N_z \ln \sigma^2$, and $\ln|\boldsymbol{\omega}^{-1}|$ are constant, and they do not affect model selection. The only differences between different models are rendered by $\frac{\boldsymbol{\varepsilon}^{*T}\boldsymbol{\omega}\boldsymbol{\varepsilon}^*}{\sigma^2}$. Here, $\frac{\boldsymbol{\varepsilon}^{*T}\boldsymbol{\omega}\boldsymbol{\varepsilon}^*}{\hat{\sigma}^2}$ is used to estimate $\frac{\boldsymbol{\varepsilon}^{*T}\boldsymbol{\omega}\boldsymbol{\varepsilon}^*}{\sigma^2}$.

Then model selection criteria *AIC* (*Akaike*, 1974) and *AICc* (*Hurvich and Tsai*, 1989) for model $M_k$ are defined as:

$$AIC_k = -2\ln[L(\hat{\boldsymbol{\beta}}_k \mid \mathbf{z}^*)] + 2N_k \tag{A6}$$

$$AICc_k = -2\ln[L(\hat{\boldsymbol{\beta}}_k \mid \mathbf{z}^*)] + 2N_k + \frac{2N_k(N_k+1)}{N_z - N_k - 1} \tag{A7}$$

Substituting (A2) into (A6) and (A7) and dropping all constants, which do not affect model selection, leads to

$$AIC_k = \frac{\boldsymbol{\varepsilon}^{*T}\boldsymbol{\omega}\boldsymbol{\varepsilon}^*}{\hat{\sigma}^2} + 2N_k \tag{A8}$$

$$AICc_k = \frac{\boldsymbol{\varepsilon}^{*T}\boldsymbol{\omega}\boldsymbol{\varepsilon}^*}{\hat{\sigma}^2} + 2N_k + \frac{2N_k(N_k+1)}{N_z - N_k - 1} \tag{A9}$$

In (A8) and (A9), $\frac{\boldsymbol{\varepsilon}^{*T}\boldsymbol{\omega}\boldsymbol{\varepsilon}^*}{\hat{\sigma}^2}$ measures the goodness of fit between estimated and observed state variables; $2N_k$ and $2N_k + \frac{2N_k(N_k+1)}{N_z - N_k - 1}$ indicate model complexity in model selection criteria.

*AIC* and $AIC_C$ values for different PTFs and ensemble models are shown in Table S2.

**Text S4 The Weighting for Field Capacity and Wilting Point in NCSS Dataset**

Because of the abudance of field capacity (pressure head at 330 cm) and wilting point (pressure head at 15,000 cm) in NCSS dataset, the weights for these two retention points were optimized for different PTF models, while the data for saturated water content were rarely observed and therefore were not analyzed in this study. Saturated water content (as well as moisture content at any arbitrary pressure head) can be predicted by evaluating the estimated BC and VG parameters.

The weights were optimized by minimizing the misfit between estimated and observed moisture content, i.e., Equation (3) in the main text. The weights of different PTFs for field capacity and wilting point are shown in Figure S1. The trends for field capacity and wilting point are generally similar, but there are some exceptions. For example, the weights of Rawls, Campbell, and Vereecken PTFs have different patterns for field capacity and wilting point. The reason for the high weight of field capacity and low weight of wilting point for Campbell PTF is probably that residual water content is assumed as 0 in Campbell PTF, which degraded that estimation of wilting point (close to residual water content). The contrary performance for field capacity and wilting point with respect to Rawls is likely that this PTF has a better capability to estimate residual water content by using a regression equation to sand, silt percentages, and bulk density, while saturated water content is derived simply from bulk density (saturated water content = 1-bulk density/2.65). The same reason may be applied to the Vereecken PTF.

**Text S5 The Variation of CV values for Saturated Water Content, Field Capacity and Wilting Point**

High CV values for saturated water content (Figure 3*b*) are found in most in high latitudes of the North Hemisphere, while high CV for field capacity (Figure 3*d*) is found in the Sahara and Arabian Peninsula (which have high sand content); South American, central Asia, and Northwest of China are shown to have high variability of the wilting point (Figure 3*f*). The CV reflects different estimates by the individual PTFs, as well as slightly different weights assigned during the optimization for each of the 100 bootstrap replicas of the selected NCSS data. We note that CV values for the full 13-member ensemble are much higher than the Group A through D ensembles (See Figures S2, S3 and S4).

**Table S1. Overview of Pedotransfer functions (PTFs) used to describe moisture retention characteristics in this study.**

| PTFs | Source | No. of citations* | Soil water retention model | No. of samples | Textural Classes | Textural Percentage | Bulk Density | Organic Carbon |
|---|---|---|---|---|---|---|---|---|
| Cosby0 | Cosby et al., (1984) | 1,343 | CMP | 1,448 | + | | | |
| Carsel | Carsel and Parrish (1988) | 1,801 | BC-VG | 5,097-5,693 | + | | | |
| Clapp | Clapp and Hornberger (1978) | 2,378 | CH | 1,446 | + | | | |
| Rosetta3-H1*w* | Zhang and Schaap (2017) | 1,638** | VG | 2,134 | + | | | |
| Cosby1 | Cosby et al., (1984) | See Cosby0 | CMP | 1,448 | | + | | |
| Cosby2 | Cosby et al., (1984) | See Cosby0 | CMP | 1,448 | | + | | |
| Rosetta3-H2*w* | Zhang and Schaap (2017) | See Rosetta3-H1*w* | VG | 2,134 | | + | | |
| Rawls | Rawls and Brakensiek (1985) | 572 | BC, CMP, And VG | 5,320 | | + | + | |
| Campbell | Campbell and Shiozawa (1992) | 237 | CMP | 6 | | + | + | |
| Rosetta3-H3*w* | Zhang and Schaap (2017) | See Rosetta3-H1*w* | VG | 2,134 | | + | + | |
| Wosten | Wösten et al., (1999) | 911 | VG | 5,521 | | + | + | + |
| Weynants | Weynants et al., (2009) | 96 | VG | 166 | | + | + | + |
| Vereecken | Vereecken et al., (1989) | 771 | VG | 182 | | + | + | + |

\* Number of citations were checked in scholar.google.com on Sept 6, 2018.

\*\* The number of citations includes the citations of Rosetta1 (Schaap et al., 2001) and Rosetta3 (Zhang and Schaap, 2017) hierarchical PTFs.

Abbreviated references, BC: Brooks and Corey (1964), BC-VG: BC parameters converted to VG parameters;, CH: Clapp and Hornberger (1978), CMP: Campbell (1974), VG: van Genuchten (1980), VG*: modfied van Genuchten (1980).

**Table S2. AIC and AICc calculated for different PTFs and ensemble models.**

| PTFs | RMSE* | $N_z$** | $N_k$** | AIC value | AICc value |
|---|---|---|---|---|---|
| Cosby0 | 0.0624 | 118599 | 1 | 106671.93 | 106671.93 |
| Carsel | 0.0987 | 118599 | 1 | 266876.53 | 266876.53 |
| Clapp | 0.0627 | 118599 | 1 | 107700.07 | 107700.07 |
| Rosetta3-H1*w* | 0.0681 | 118599 | 1 | 127049.77 | 127049.77 |
| Cosby1 | 0.0607 | 118599 | 1 | 100938.96 | 100938.96 |
| Cosby2 | 0.062 | 118599 | 1 | 105308.75 | 105308.75 |
| Rosetta3-H2*w* | 0.0628 | 118599 | 1 | 108043.88 | 108043.88 |
| Rawls | 0.0629 | 118599 | 1 | 108388.23 | 108388.23 |
| Campbell | 0.0675 | 118599 | 1 | 124820.91 | 124820.91 |
| Rosetta3-H3*w* | 0.0589 | 118599 | 1 | 95041.34 | 95041.34 |
| Wosten | 0.0565 | 118599 | 1 | 87454.00 | 87454.00 |
| Weynants | 0.0555 | 118599 | 1 | 84385.74 | 84385.74 |
| Vereecken | 0.0658 | 118599 | 1 | 118612.90 | 118612.90 |
| Ensemble: Group A | 0.0601 | 118599 | 4 | 98959.36 | 98959.36 |
| Ensemble: Group B | 0.0589 | 118599 | 3 | 95045.34 | 95045.34 |
| Ensemble: Group C | 0.0536 | 118599 | 3 | 78711.01 | 78711.02 |
| Ensemble: Group D | 0.0528 | 118599 | 3 | 76379.14 | 76379.14 |
| Ensemble: Overall | 0.0517 | 118599 | 13 | 73250.08 | 73250.08 |
| Stratification of Soil Textural Classes | 0.0511 | 118599 | 156 | 71846.35 | 71846.76 |
| Stratification of Soil OC | 0.0511 | 118597 | 104 | 71741.14 | 71741.33 |
| Stratification of Soil Order | 0.0506 | 91303 | 156 | 54310.03 | 54310.56 |
| Stratification of Soil Temperature | 0.049 | 80223 | 78 | 44648.10 | 44648.25 |

* RMSE value is from Table 1. ** $N_z$ is the number of observed water content; $N_k$ is the number of parameters used in optimization.

**Table S3.** *RMSE* values of calibration and validation results of ensemble models using NCSS dataset and weights of all PTFs, including Cosby0 (lookup table by Cosby et al. 1984), Carsel (Carsel and Parrish 1988), Clapp (Clapp and Hornberger, 1978), Cosby1 (one way analysis of variance by Cosby et al. 1984), Cosby2 (two way analysis of variance by Cosby et al. 1984), Rawls (Rawls and Brakenssiek, 1985), Campbell (Campbell and Shiozawa, 1992), Wosten (Wosten et al. 1999), and Weynants (Weynants et al. 2009), Vereecken (Vereecken et al. 1989), and Rosetta3 - H1w, H2w, H3w models (Zhang and Schaap 2017).

**Table S4.** *RMSE* values of calibration and validation results of ensemble models using NCSS dataset for PTFs of Group A (using USDA textural classes as predictors) and weights of corresponding PTFs. Group A includes Cosby0, Carsel, Clapp, and Rosetta3 - H1w PTFs.

**Table S5.** *RMSE* values of calibration and validation results of ensemble models using NCSS dataset for PTFs of Group B (using soil textural percentages as predictors) and weights of corresponding PTFs. Group B includes Cosby1, Cosby2, and Rosetta3 – H2w PTFs.

**Table S6.** *RMSE* values of calibration and validation results of ensemble models using NCSS dataset for PTFs of Group C (using soil textural percentages and bulk density as predictors) and weights of corresponding PTFs. Group C includes Rawls, Campbell, and Rosetta3 – H3w PTFs.

**Table S7.** *RMSE* values of calibration and validation results of ensemble models using NCSS dataset for PTFs of Group D (using soil textural percentages, bulk density and organic carbon as predictors) and weights of corresponding PTFs. Group D includes Rawls, Campbell, and Rosetta3 – H3w PTFs.

**Table S8.** *RMSE* values of calibration and validation results of ensemble models for 12 USDA soil textural classes and weights of corresponding PTFs. 12 USDA soil textural classes includes Sa, sand; LoSa, loamy sand; SaLo, sandy loam; SaClLo, sandy clay loam; SaCl, sandy clay; Lo, loam; SiLo, silty loam; Si, silt; SiClLo, silty clay loam; SiCl, silty clay; ClLo, clay loam; Cl, clay.

**Table S9.** *RMSE* values of calibration and validation results of ensemble models for eight different soil organic carbon content and weights of corresponding PTFs.

**Table S10.** *RMSE* values of calibration and validation results of ensemble models for six types of soil temperature and weights of corresponding PTFs. The soil temperature classifications include: frigid, hyperthermic, isohyperthermic, isomesic, mesic, and thermic.

**Table S11**. *RMSE* values of calibration and validation results of ensemble models for 12 types of soil orders and weights of corresponding PTFs. Soil orders include alfisols, andisols, aridisols, entisols, gelisols, histosols, inceptisols, mollisols, oxisols, spodosols, ultisols, and vertisols.

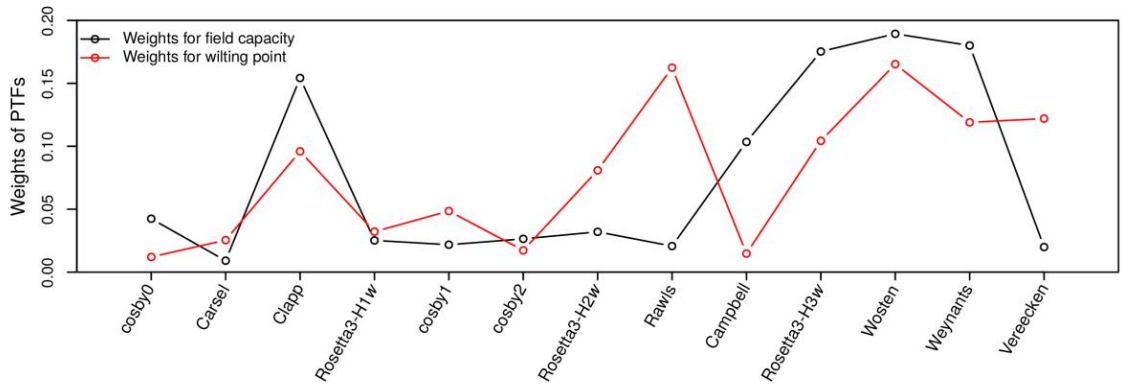

**Figure S1. Weights of PTFs versus different PTF models. Black and red lines indicate weights for field capacity and wilting point, respectively.**

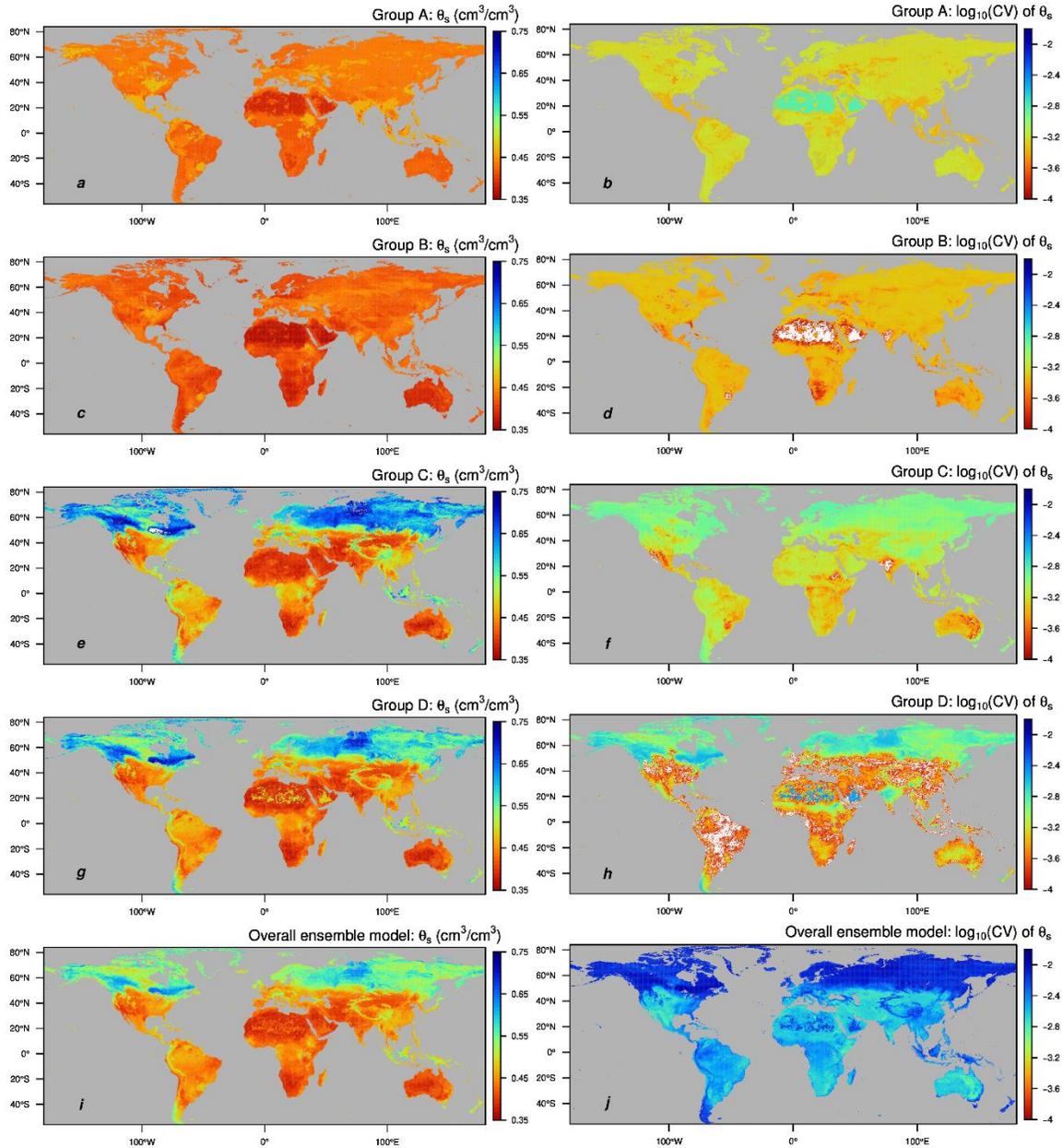

**Figure S2. Global maps of average values of saturated water content (*a*, *c*, *e*, *g* and *i*) and coefficient of variation (*b*, *d*, *f*, *h*, and *j*; using $\log_{10}$ scale for display reason) in 10 km resolution estimated from different groups of ensemble model.** White colors indicate the values are out of the range of legend. Estimations are calculated based on 100 bootstrap replicas of optimized weights. Calculations are based on the surface soil of SoilGrids 10 km data set (Hengl et al., 2014).

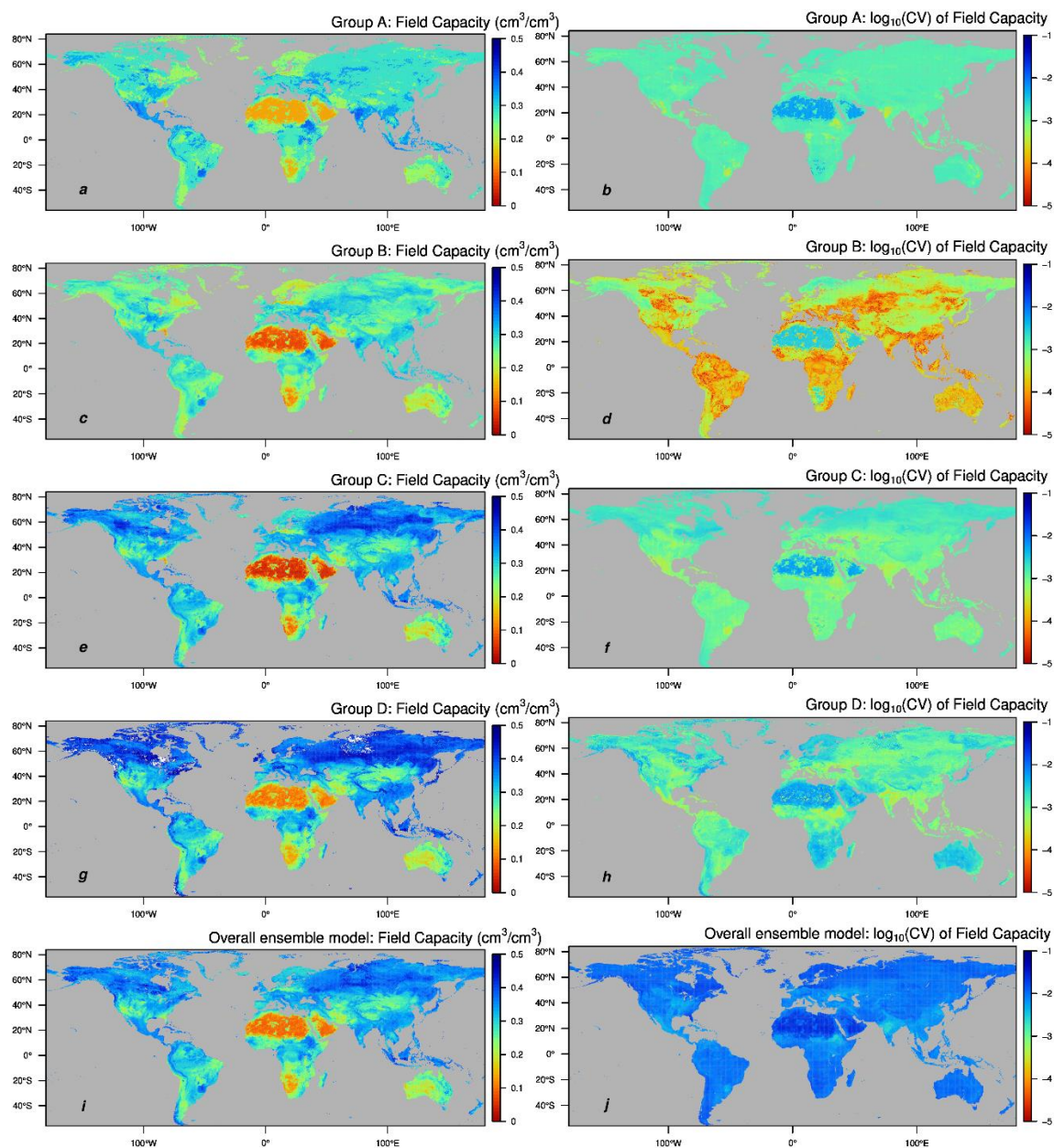

**Figure S3. Global maps of field capacity ($h$ = 330 cm) in 10 km resolution estimated from different groups of ensemble model.** *a*, *c*, *e*, *g* and *i* are maps of mean values, *b*, *d*, *f*, *h* and *j* for **coefficient of variation (abbrivated as CV; using $\log_{10}$ scale for display reason).** White colors indicate the values are out of the range of legend. Estimations are calculated based on 100 bootstrap replicas of optimized weights. Calculations are based on the surface soil of SoilGrids 10 km data set (Hengl et al., 2014).

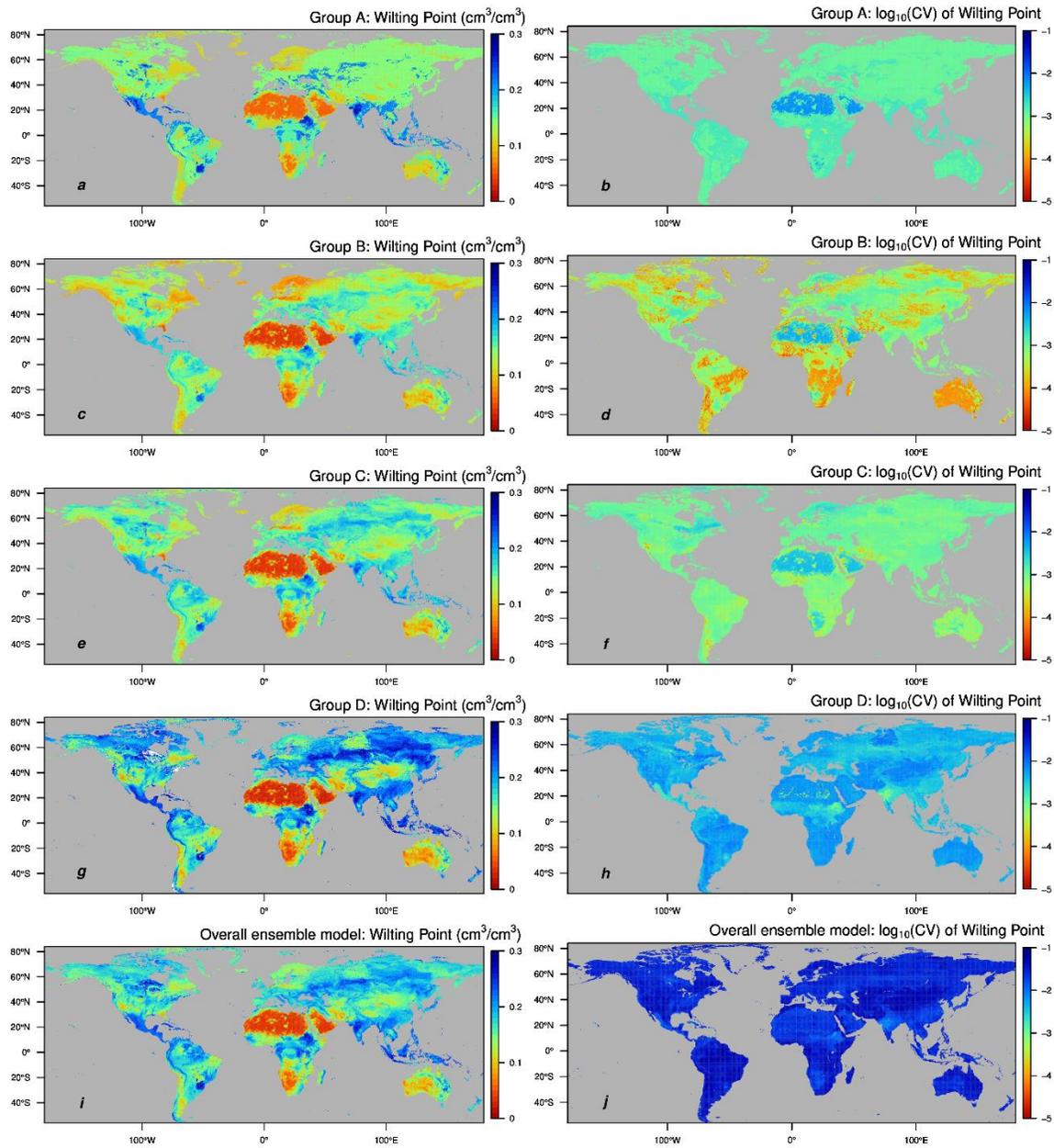

**Figure S4. Global maps of permanent wilting point ($h = 15,000$ cm) in 10 km resolution estimated from different groups of ensemble model.** *a*, *c*, *e*, *g* and *i* are maps of mean values, *b*, *d*, *f*, *h* and *j* for coefficient of variation (abbrivated as CV; using $\log_{10}$ scale for display reason). White colors indicate the values are out of the range of legend. Estimations are calculated based on 100 bootstrap replicas of optimized weights. Calculations are based on the surface soil of SoilGrids 10 km data set (Hengl et al., 2014).